\documentclass[aip,rsi,reprint]{revtex4-1}

\usepackage{graphicx}

\begin{document}


\title{Temperature-stabilized differential amplifier for low-noise DC measurements}




\author{P. M\"{a}rki}
\email[]{pmaerki@phys.ethz.ch}
\affiliation{ETH Z\"urich, Solid State Physics Laboratory, Otto-Stern-Weg 1, 8093 Z\"urich, Switzerland}
\author{B. A. Braem}
\affiliation{ETH Z\"urich, Solid State Physics Laboratory, Otto-Stern-Weg 1, 8093 Z\"urich, Switzerland}
\author{T. Ihn}
\affiliation{ETH Z\"urich, Solid State Physics Laboratory, Otto-Stern-Weg 1, 8093 Z\"urich, Switzerland}


\date{\today}

\begin{abstract}
A tabletop low-noise differential amplifier with a bandwidth of 100\,kHz is presented. Low voltage drifts of the order of 100\,nV/day are reached by thermally stabilizing relevant amplifier components. The input leakage current is below 100\,fA. Input-stage errors are reduced by extensive circuitry. Voltage noise, current noise, input capacitance and input current are extraordinarily low. The input resistance is larger than 1\,T$\Omega$. The amplifiers were tested with and deployed for electrical transport measurements of quantum devices at cryogenic temperatures.

\end{abstract}

\pacs{}

\maketitle 

\section{Introduction}
Many scientific fields require low-noise dc amplifiers. For example, in cryogenic experiments samples are cooled below 100\,mK to measure their electrical properties which are governed by quantum effects and quantum coherence. In these devices and at these low temperatures, the voltages to be measured and also the noise generated by the samples themselves are tiny: a sample with a resistance of 1\,M$\Omega$ shows a Johnson-voltage noise below 1\,nV$/\sqrt{\mathrm{Hz}}$ at 10\,mK. The dominant measurement noise is therefore caused by room-temperature amplifiers. Lock-in techniques can be used for improving the signal-to-noise ratio. However, these methods are problematic, if long cables and filters with significant stray capacitances have to be employed between amplifier and sample for avoiding heating and decoherence of the device electrons. Phase-shifts, that depend on the (varying) sample impedance can completely distort the signals acquired with the lock-in amplifier.

Another approach towards low noise measurements is cooling the first amplifier stages at cryogenic temperatures. \cite{2006, 2011} The price to pay is a massively higher complexity of the resulting measurement system and the loss of flexibility when performing experiments.
This is only one of many examples, where ultra-low noise room-temperature dc amplifiers for the detection of low-level voltages are advantageous. 

Amplifiers with an input voltage-noise in the range of 1\,nV$/\sqrt{\mathrm{Hz}}$ can be implemented with silicon junction field-effect transistors (FETs). Low input currents as low as a few \,pA can be achieved, but the leakage current typically doubles every 10$^{\circ}$C temperature increase. At low source resistances the leakage current of the amplifier is not an issue. However, if the source resistance is larger than a few k$\Omega$, the voltage drop across it caused by the leakage current modifies the measurement result significantly.

In this paper we present a tabletop dc amplifier design for low-level voltages that minimizes the low-frequency noise (4.7\,nV$_{\mathrm{rms}}$ 0.1\,Hz to 10\,Hz) in combination with an input DC current below 100\,fA, a very remarkable figure of merit. Stabilizing the temperature of the FET, we keep the leakage current constant such that it can be compensated with a constant current source. With this principle the resulting leakage current of the amplifier input can be reduced significantly. Moreover the constant temperature reduces the $1/f$ input voltage-noise caused by random temperature fluctuations. Together with a suitable design of the input stage this leads to a very high input resistance, high common-mode rejection and low input capacitance.

\begin{figure}
 \includegraphics[width=\linewidth]{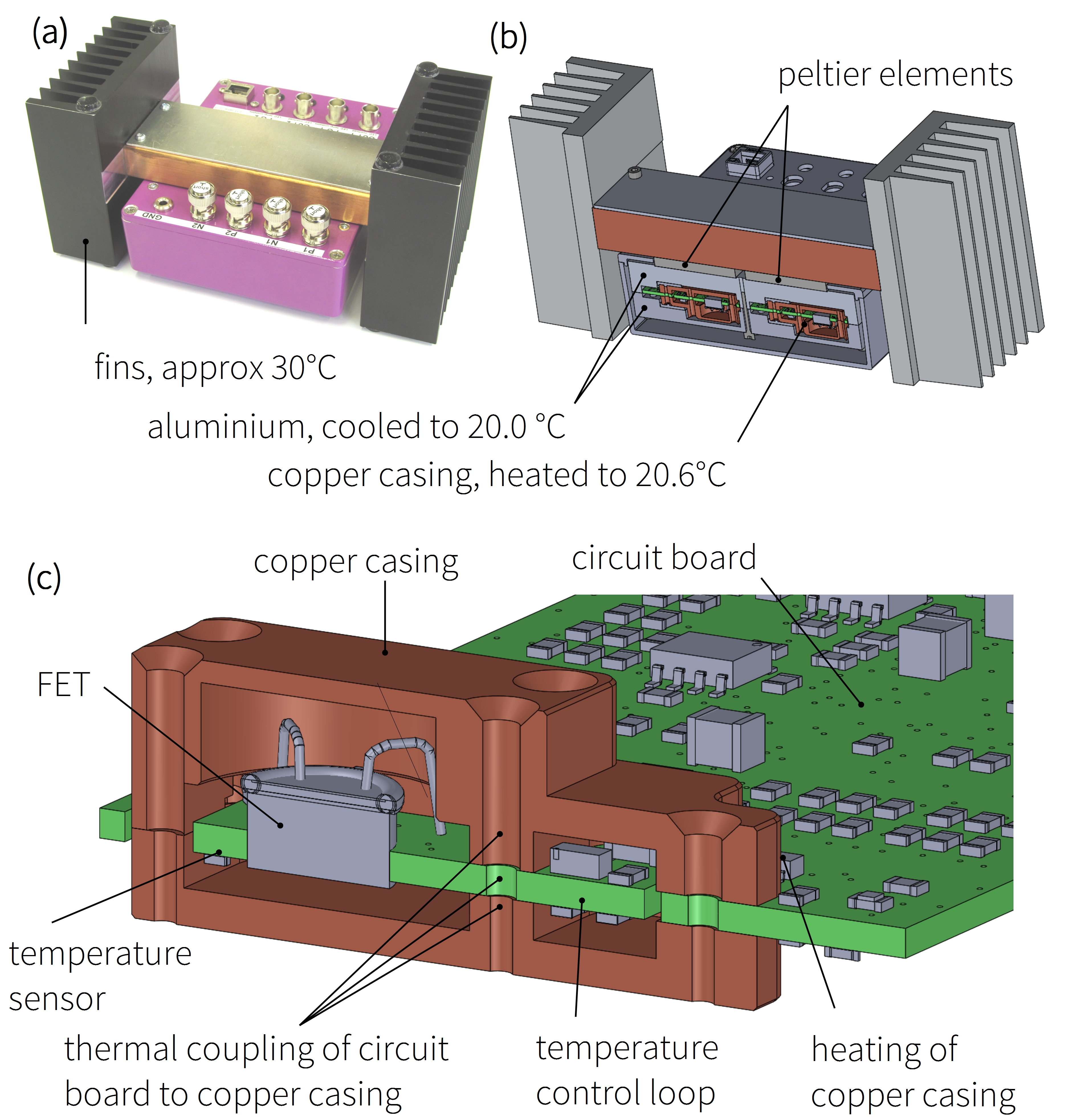}
 \caption{(a) The amplifier consists of two independent differential channels. (b) Cross section through amplifier. (c) Cross cut through circuit board and copper casing.}
 \label{grafik_temperatur}
\end{figure}

\section{thermal stabilization}

The design and realization of the temperature stabilisation is presented in Fig.~\ref{grafik_temperatur}. The circuit-board is enclosed within an aluminium block cooled to 20$^{\circ}$C by Peltier elements [see Fig.~\ref{grafik_temperatur}(b)]. A massive copper bar guides the excess heat to the heat fins seen in Figs.~\ref{grafik_temperatur}(a) and (b). The gap between circuit board and aluminium is filled with soft heat-conductive gap filler to improve thermal coupling and to avoid airflow. The input-stage double-FET is placed inside a copper casing stabilized at 20.6$^{\circ}$C [Figs.~\ref{grafik_temperatur}(b) and (c)] by an analog temperature control loop. The double FET casing is soldered onto a small peninsula of the circuit board [Figs. \ref{leckstrom}(b) and (c)] to optimize thermal anchoring of the FETs. The peninsula reduces mechanical strain on the FET casing and wires. It is surrounded but not touched by the copper casing. The air volume around the FET is small to avoid thermal fluctuations due to convection. The temperature fluctuation of the peninsula was measured to be below 1\,mK during one day. The circuit board is clamped between the upper and the lower copper part for thermalizing the striplines entering the peninsula. Electrical insulation is achieved with a thin polyimide tape.

The choice of temperatures around 20$^{\circ}$C  reflects the compromise between avoiding water condensation at high environmental humidity and increased leakage currents at higher temperatures. Besides this, minimizing the temperature difference between amplifier and external cabling reduces undesired thermoelectric voltages.

The temperature inside the semiconductor of the FET can change, if the dissipated power in the FET is modulated. Since it is not possible to measure or control this change with a sensor outside the FET casing, it is essential to keep the FET's power loss constant. This is achieved with an appropriate circuit design (see below). After powering up the amplifier it takes about 20 minutes until all temperatures are stable.

We found experimentally that with the temperature stabilization in operation, amplifier drifts were almost completely suppressed, while they reappeared once the temperature control loop was switched off.
Similar approaches of stabilizing the temperature are widely used, for example, in oven controlled crystal oscillators or precision voltage references.\cite{ocxc} Further reduction of the FET temperature reduces leakage currents to negligible values.\cite{1996}

\section{first amplification stage}
\begin{figure}
 \includegraphics[width=\linewidth]{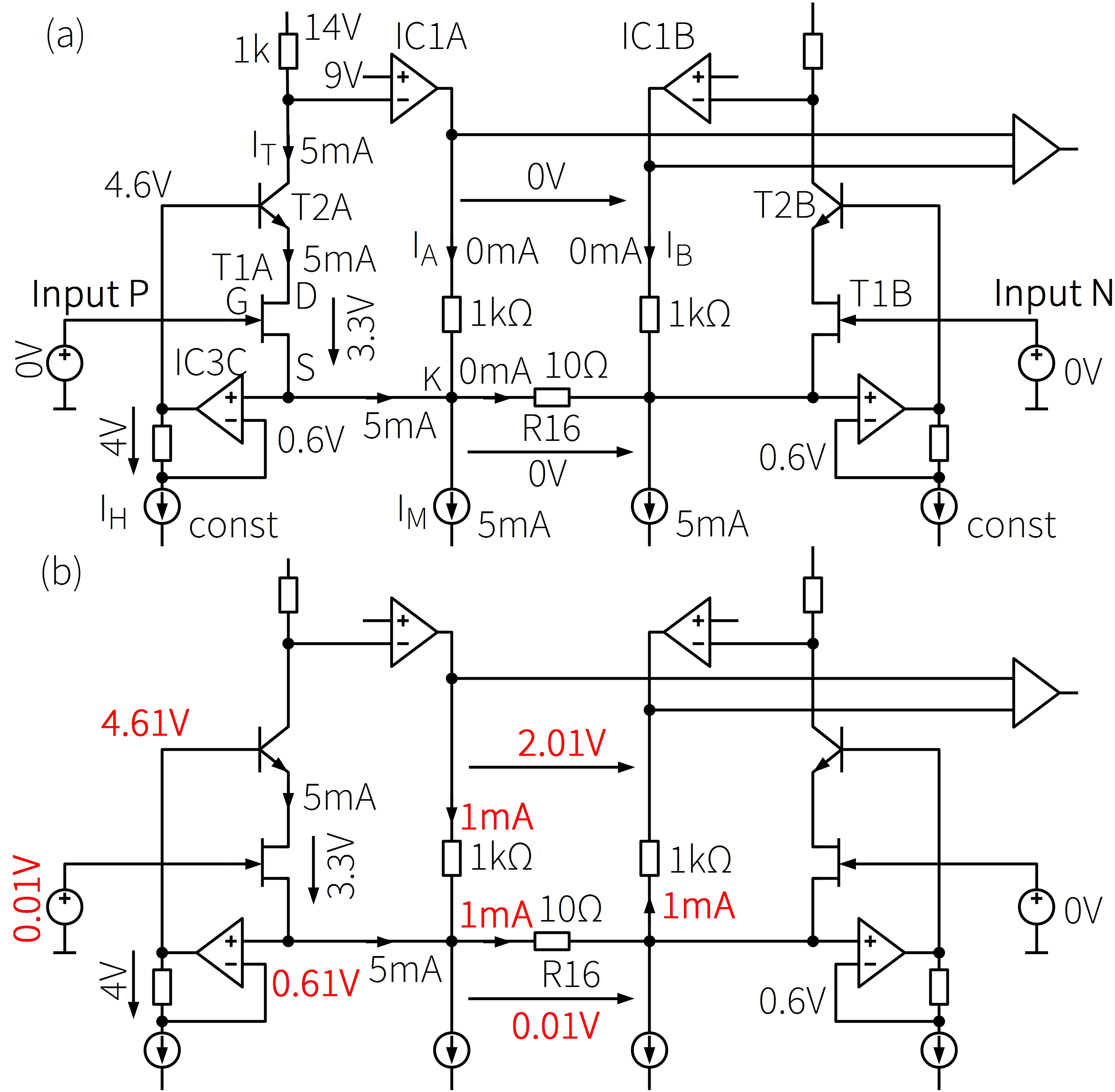}
 \caption{Simplified schematic of the first differential amplification stage and its principle of operation. The FETs operate at a fixed working-point with 3.3\,V drain-source voltage and 5\,mA drain current. As a consequence, the gate-source voltages are constant as well. (a) Voltages of 0\,V are applied at both inputs P and N. The voltage drop across R16 is zero by symmetry. (b) A voltage of 10\,mV is applied at Input P, while N remains at zero. Both the source of T1A and the base of T2A are shifted by 10\,mV, but the drain source voltage remains at 3.3\,V and the drain current at 5\,mA. The resulting voltage across R16 leads to a transverse current of 1\,mA. The generated voltage of 2.01\,V is further amplified by the subsequent stages of the amplifier (not shown).}
 \label{schematic}
\end{figure}

Figure~\ref{schematic}(a) shows the symmetric circuit design of the amplifier input stage highlighting its basic functionality and omitting the leakage current compensation to be discussed later.
It is built around the FETs T1A and T1B. 

In the balanced situation, where zero volts are applied to both inputs, the voltage drop across R16 is zero by symmetry. Therefore the currents $I_{\mathrm{A}}$ and $I_{\mathrm{B}}$ are zero as well and the difference between the outputs of IC1A and IC1B is $0\,\mathrm{V}$.

Let us first discuss how IC1A forces a constant current $I_{\mathrm{T}}$ trough the transistor branch. The non-inverting input of IC1A is set at a constant voltage of $9\,\mathrm{V}$. The inverting input of IC1A is at $9\,\mathrm{V}  =  14\,\mathrm{V} - 5\,\mathrm{mA}\times1\,\mathrm{k}\Omega$ as a consequence of the feedback. If $I_{\mathrm{T}}$ deviates from  $5\,\mathrm{mA}$ the output of IC1A will force a current $I_{\mathrm{A}}$ into node~K until $I_{\mathrm{T}}$ reaches $5\,\mathrm{mA}$ again.
The constant current $I_{\mathrm{T}}$ = $5\,\mathrm{mA}$ flows into node~K and further into the constant current sink I$_{\mathrm{M}}$. This, in turn, implies that any current flowing through R16 will be equal to $I_{\mathrm{A}}$ and supplied by IC1A.

Next, we explain the function of IC3C and the constant-current source I$_{\mathrm{H}}$. This combination is such that the output of IC3C is 4\,V above the source voltage of T1A, thereby bootstrapping\cite{bootstrapping} the base of T2A and the source of T1A at a constant value. Due to the fact that the base--emitter voltage of T1A is practically constant at 0.7\,V, the drain--source voltage of T1A is kept at a constant value of 3.3\,V independent of collector--emitter voltage.

The two properties of the circuit discussed above, the constant T1A current of 5 mA and the constant T1A drain--source voltage of 3.3 V, lead to constant power dissipation in the FET and therefore keep the internal temperature of the FET constant, independent of the applied input voltage. They are the key to the superior stability of the amplifier, even at high input voltages.~\cite{largestepcoment} Furthermore, they reduce the parasitic input capacitance by bootstrapping the drain-gate capacitance.

Figure~\ref{schematic}(b) highlights the changes of the voltages and currents in the circuit, once the input voltages at P and N differ. As described before, the current through T1A and the drain--source voltage across T1A have to remain the same. A side effect is that the gate--drain voltage at T1A will also remain the same. The new input voltage on input P therefore causes a voltage imbalance across R16 resulting in a current given by the input voltage difference divided by the resistance of $10\,\Omega$. This current has to flow between IC1A and IC1B causing a voltage difference between the two outputs which is 201 times bigger than the input voltage difference. As a consequence, the gain of the input stage of the amplifier is $2\times 1\,\mathrm{k}\Omega/10\,\Omega + 1 = 201$. A second amplification stage amplifies the output voltage of the first stage to a total gain of 1000. As the second stage does not contribute significantly to the noise level, it is not discussed here.

The Johnson-noise contribution of each component in the first amplification stage can be calculated and is well understood.~\cite{scandurra} In contrast, low-frequency noise is harder to predict and its sources are difficult to localize. Nevertheless, this is crucial for the performance of DC-amplifiers. Low-frequency  noise contributions stem from flicker noise mainly generated in the semiconductor material of the FET. In addition, there are thermoelectric voltage drifts caused by temperature variations, resistor flicker noise and others. The direct measurement of such single noise sources is very difficult, because temperature drifts dominate when the amplifier casing is left open. Therefore the effects of a modification cannot be measured before reassembling and temperature stabilizing the device.

\section{leakage current}
\begin{figure}
 \includegraphics[width=\linewidth]{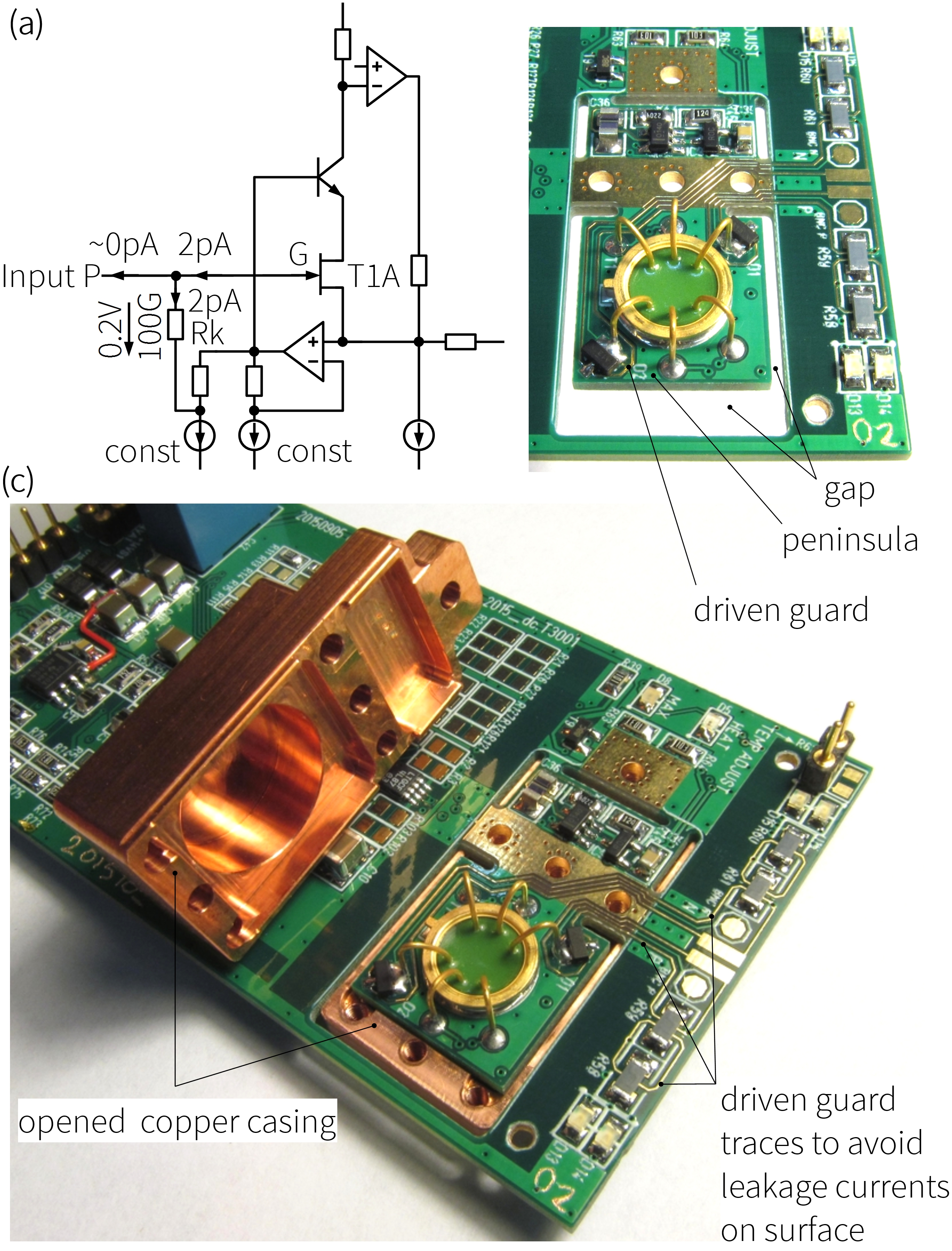}
 \caption{(a) The left half of the schematic of Fig. \ref{schematic}(a)  with additional leakage compensation. The current source built with Rk is adjusted to compensate the leakage current of the FET. In the example a leakage current of 2\,pA is flowing out of the gate of the FET T1A. The same current is compensated by a current source built with a 100\,G$\Omega$ resistor. Only the difference of these currents flows out or into the input P. (b) Circuit board without copper casing. The peninsula shape reduces mechanical stress to the circuit board and the sensitive FET. Mechanical stress can influence the electrical properties of the FET and induce charges due to piezoelectricity. (c) View on the circuit board, copper casing opened. All sensitive traces are enclosed by driven guard traces (gold).}
 \label{leckstrom}
\end{figure}

The input leakage current of the FET is constant because the temperature of the FET is kept constant. It is compensated by a constant current source as shown in Fig.~\ref{leckstrom}(a). Typical gate currents are in the range of a few pA. Even though the leakage current can be compensated, it is important to choose a FET and operating point with a low value, because the gate current and the compensation current increase the current noise at the amplifier input. A low operating temperature, low $V_{\mathrm{DS}}$ and low drain current help to keep the gate leakage low. The drawback of low $V_{\mathrm{DS}}$ and low drain current is the reduced transconductance of the FET and, as a consequence, the higher voltage noise. Additionally, all sensitive traces are enclosed by driven guard~\cite{driven_guard, bootstrapping} traces (gold) to reduce leakage currents on the circuit board surface, as seen in Figs.~\ref{leckstrom}(b) and (c). There are more driven guard traces on the inner circuit board layers.

\section{Design of the amplifier power supply}
Time-varying magnetic stray-fields of transformers can disturb small signals by induction. Therefore a power-supply with a toroidal transformer in a separate casing is used to power the amplifier. The power-supply should be placed at a distance of more than 2$\,\mathrm{m}$ from the amplifier and the signal-generating circuits. The linear stabilized $\pm15\,\mathrm{V}$ of the power supply are further reduced to $\pm14\,\mathrm{V}$ in the amplifier. Here a voltage reference followed by low pass filters is used to get a low-noise supply voltage. 
The AC ground leakage current (power line frequency and harmonics) of the power supply flows via the measurement cables to the ground of the experiment and can therefore disturb the measurement signal. A very low leakage current below 300\,nA could be realized by special shielding techniques in the custom made toroidal transformer.
The casing of the power supply also hosts the analog control loop for the Peltier cooling system. This design keeps the dissipated power of the Peltier current controller remote from the thermally stabilized amplifier.

\section{thermoelectric voltages}
Thermoelectric voltages can exceed the measured voltages by orders of magnitude. Therefore special attention has been paid to avoid temperature differences and the combination of materials developing strong thermoelectricity.
The BNC connectors used at the inputs are not optimal in this respect, but standard in our laboratories. When a person touches a connector, a thermoelectric voltage of more than 1\,$\mu$V can be observed due to the thermoelectric voltages occuring between the different materials. Silicon rubber hoses (7 cm long) are pushed over the plugs at the ends of the BNC cables to reduce variations in their temperatures.

\section{selecting components}
It is very difficult to locate the sources of low-frequency noise in a low-noise amplifier. A single bad component can spoil the noise performance. It is therefore essential to test the components prior to assembly: the input FET, some active components and the leakage current of some capacitors. Only a fraction of the components fulfils the requirements and can be used. Such measurements take a long time, in particular for identifying the presence of bursts of telegraph noise which can very rarely occur in FETs. 
However, given the time, effort, and financial resources invested in cutting-edge research, this preselection process is justified.

\section{adjustment}
The circuitry has to be adjusted to get low offset voltage, low input current and a good common-mode rejection ratio. Potentiometers are not sufficiently stable in the long term and show a lot of undesired flicker-noise. Therefore some resistance values are adjusted by combining a number of fixed-value thin-film resistors instead. A description of the iterative process is provided in the supplemental material.

\section{input protection}
A dedicated circuit protects the first amplification stage: A junction FET with an impedance of 5\,$\Omega$ is connected in series to each input. At a voltage drop of about 10 mV over the FET the protection circuit starts to oscillate. The alternating signal is amplified by a transformer and then rectified. This signal pinches off the FET. Together with consecutive protection diodes, this circuit protects the sensitive components against over-voltages up to $\pm20\,\mathrm{V}$ at the input. This is required to protect the device against operation errors and technical failures: For our transport experiments we use gate voltages of several volts in close proximity to the connections of the amplifier inputs. Without input protection a flashover or a connection error might destroy the input stage of the amplifier.

\section{only analog circuits}
No digital components are used in the entire amplifier. All temperature control loops are analog. This also applies to the power supply and the controller for the Peltier elements. Therefore no high frequency switching noise will appear at the inputs. This is important in cryogenic experiments on semiconductor nanostructures to reach the lowest possible electron temperatures.

\section{characteristic figures}
\begin{figure}
 \includegraphics[width=\linewidth]{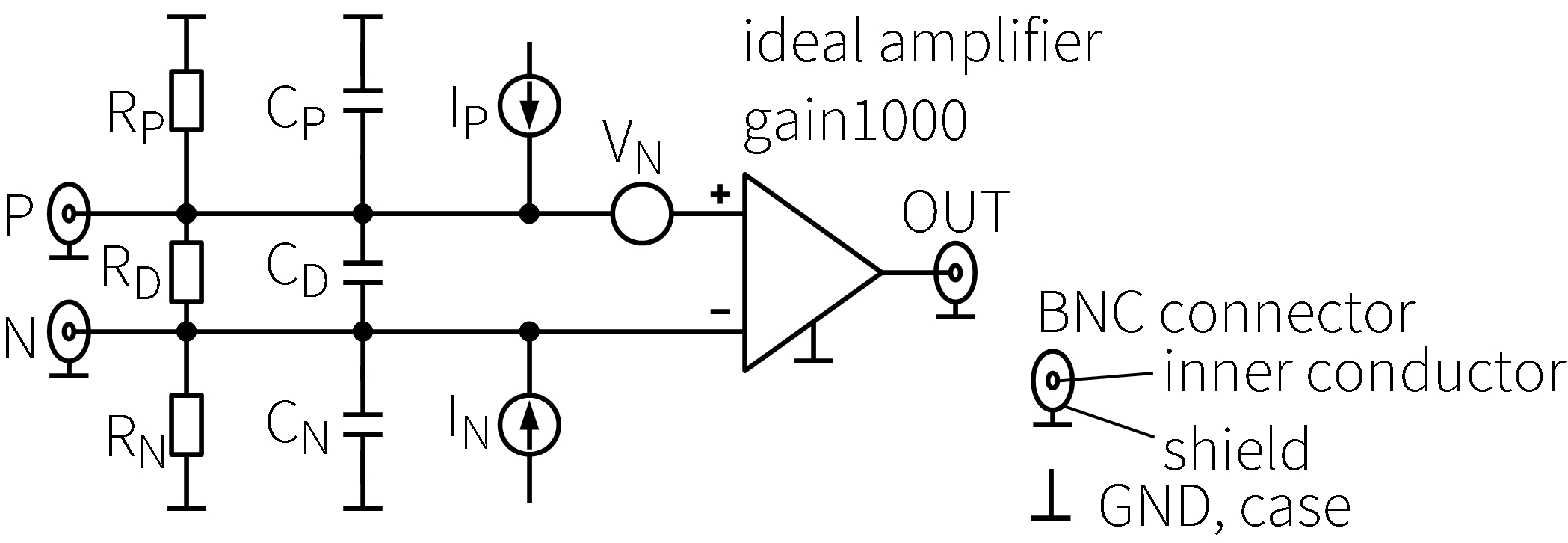}
 \caption{Parasitic model: resistors, capacitors, current sources and voltage sources are parasitic. The amplifier in this schematic is ideal.}
 \label{grafik_fehlermodell}
\end{figure}

Figure \ref{grafik_fehlermodell} shows a simple parasitic value model. The numbers and measurements shown below are typical for the first three prototypes investigated. 

\begin{figure}
 \includegraphics[width=\linewidth]{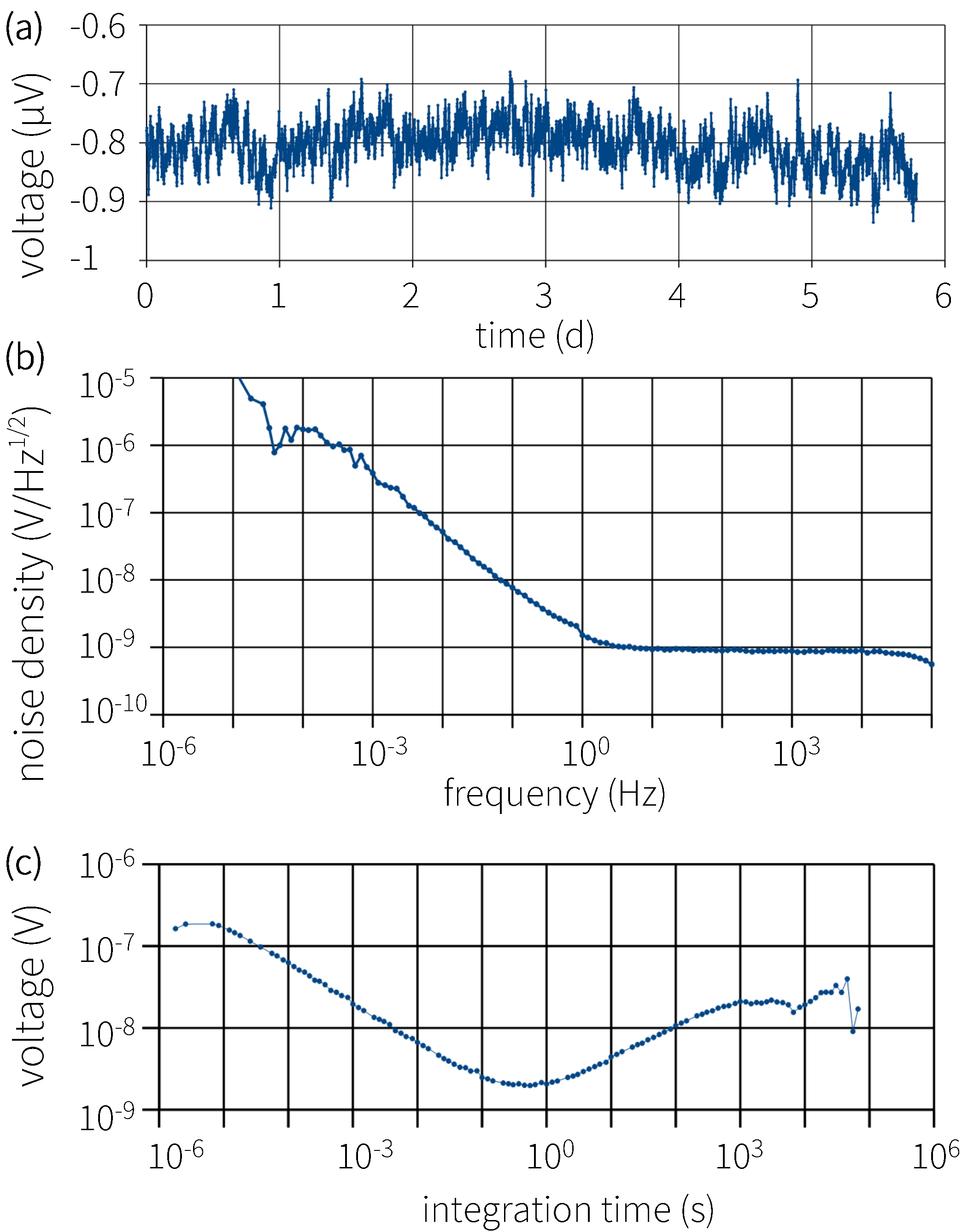}
 \caption{Input voltage V$_{\mathrm{N}}$ according to model in Fig. \ref{grafik_fehlermodell}, inputs shorted. (a) Voltage versus time, integration time 100\,s. (b) Voltage noise density. The graphic is a combination of several analysed time-traces to reduce the amount of stored data. (c) Allan Deviation. The graphic is a combination of several analysed time-traces to reduce the amount of processed data.}
 \label{grafik_spannungsrauschen}
\end{figure}

Figure \ref{grafik_spannungsrauschen}(a) shows the noise-voltage $V_\mathrm{N}$ measured with shorted inputs over a time period of 6 days. During this time the amplifier was located on the laboratory table. People walked around in the lab, windows were opened and closed. The offset-voltage remained within a band of 200\,nV testifying the superior stability of the amplifier.
Drifts below microvolts are usually achieved only by amplifiers employing auto-zeroing techniques with the drawback of chopping noise and high leakage currents.

In Fig.~\ref{grafik_spannungsrauschen}(b) we show the noise density of $V_\mathrm{N}$. The low drift is reflected by the low $1/f$-noise. Surprisingly the noise increases slightly less at frequencies below $10^{-3}\,\mathrm{Hz}$ than in the range between $1\,\mathrm{Hz}$ and $10^{-3}\,\mathrm{Hz}$. 

The Allan deviation shown in Fig.~\ref{grafik_spannungsrauschen}(c) is very useful for estimating the optimum integration time for an experiment. If, for example, the integration time is 1\,s (a voltmeter averages the signal over a time of 1s resulting in one data point every 1s), a point to point deviation of 3\,nV can be expected (Allan deviation 2\,nV @ 1\,s, for details see Refs.~\onlinecite{Allan_1966, Allan_1987})

Furthermore, the typical offset voltage drift over the time of a full measurement series can easily be estimated. If the series takes 17\,minutes, for example, a typical voltage drift of 28\,nV can be expected (Allan deviation 20\,nV @ 1000s).

\begin{figure}
 \includegraphics[width=\linewidth]{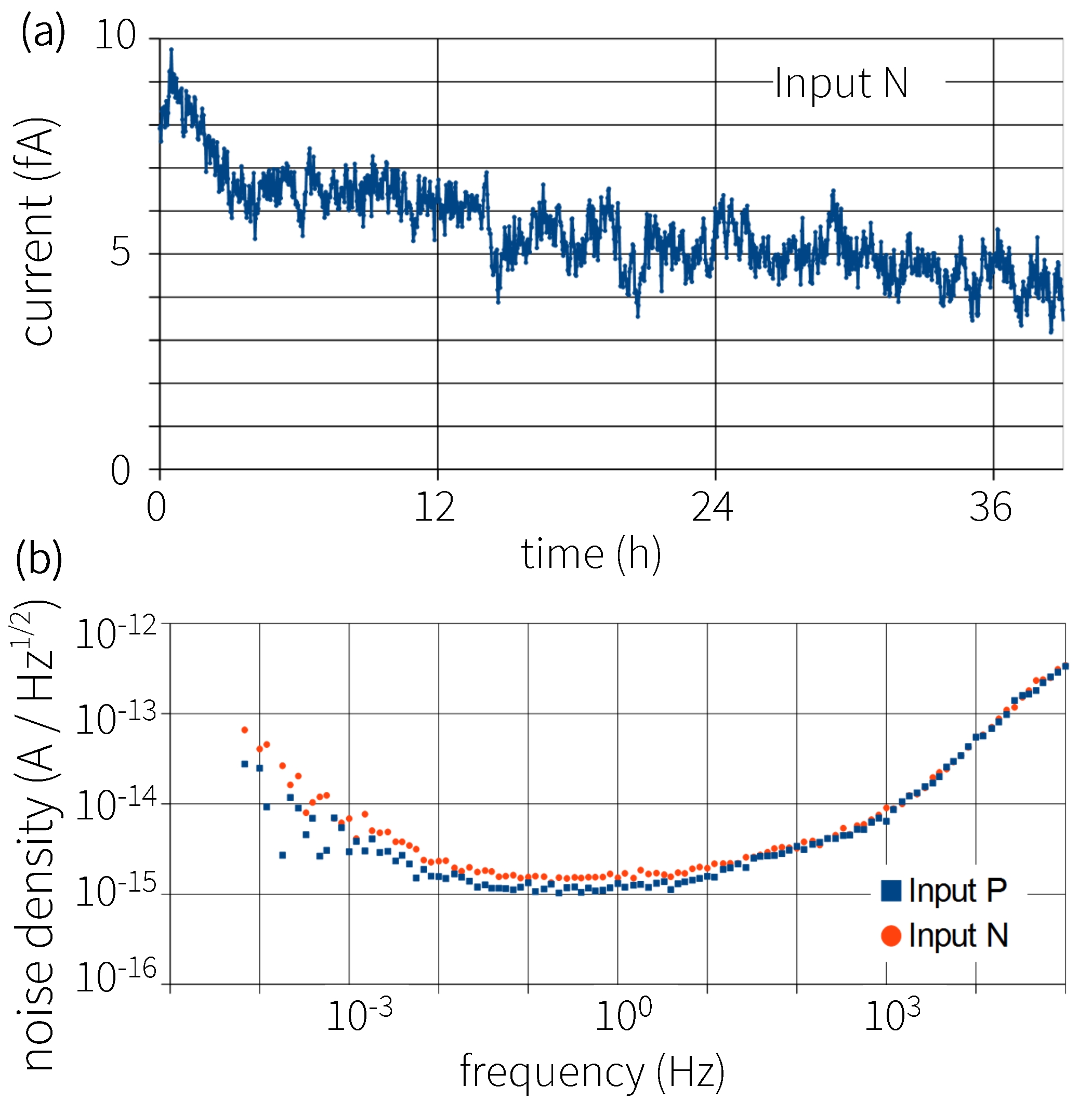}
 \caption{(a) Input leakage current I$_{\mathrm{N}}$ versus time. Integration time 100 s. (b) Current noise density of I$_{\mathrm{N}}$ and I$_{\mathrm{P}}$.}
 \label{grafik_stromrauschen}
\end{figure}

In Fig.~\ref{grafik_stromrauschen}(a) we plot the input-current $I_{\mathrm{N}}$ versus time. The current changes by a few femto Amp\`{e}re within a time span of 36\,h. The slow decrease is probably due to a slow change in humidity which can influence leakage currents on surfaces.  For this measurement a 100\,G$\Omega$ resistor was connected from input N to GND and the output voltage was measured.

The current-noise density in Fig.~\ref{grafik_stromrauschen}(b) reflects the shot noise of the gate leakage current. The slight difference between input P and input N is due to the different input leakage currents of the two FETs. The noise density increases towards higher frequencies due to capacitive coupling of the drain--source channel to the gate.

To measure the current noise, a 100\,G$\Omega$ resistor was connected between one input and ground in parallel to a 100\,pF capacitor. The total input noise is the current-noise times the input impedance plus the voltage-noise of the amplifier. The current noise consists of the Johnson current-noise of the 100\,G$\Omega$ resistor (0.4\,fA$/\sqrt{\mathrm{Hz}}$) and the input current-noise of the
amplifier. The input current-noise of the amplifier was then calculated from the measured voltage noise. The increase of current-noise towards higher frequency is due to capacitive coupling of the drain-source channel  to the gate in the first input FET.

Typical amplifier characteristics are summarized in table \ref{tabellecaracteristics}. The parasitic resistances are very high, the capacitance values are very low. This reflects the bootstrapping of the input FET and the careful layout technique. The low capacitance values are normally not important because in a typical experiment the cabling connected to the input of the amplifier has a much higher capacitance.
Voltage and current noise is very low, as shown above. The bandwidth is less important as we typically use these amplifiers below 100\,Hz.

To benefit from these characteristics it is essential to build the entire experimental set-up very carefully: cabling, shielding, temperatures, materials, measurement circuits, vibrations and so forth have to be optimized. This is often a major challenge.

\begin{table}
\caption{\label{tabellecaracteristics}Typical characteristics, parasitic model according to Fig. \ref{grafik_fehlermodell}, related to the input.}
\begin{tabular}{ l l r }
\hline
Resistance & R$_{\mathrm{P}}$, R$_{\mathrm{N}}$ & $>$10\,T$\Omega$\\
Capacitance\footnotemark[1] & C$_{\mathrm{P}}$, C$_{\mathrm{N}}$ & 10 pF\\
Differential resistance & R$_{\mathrm{D}}$ & $>$1\,T$\Omega$\\
Differential capacitance\footnotemark[1] & C$_{\mathrm{D}}$ & $<$ 1\,pF\\
Leakage current & I$_{\mathrm{P}}$, I$_{\mathrm{N}}$ & $<$ 100\,fA\\
Leakage current drift & I$_{\mathrm{P}}$, I$_{\mathrm{N}}$ & 20\,fA per day\\
Current noise density @ 1\,Hz & I$_{\mathrm{P}}$, I$_{\mathrm{N}}$ & 1.7\,fA$/\sqrt[]{\mathrm{Hz}}$\\
Offset voltage & V$_{\mathrm{N}}$ & below 2\,$\mu$V \\
Offset voltage drift & V$_{\mathrm{N}}$ & 100\,nV per day \\
Voltage noise density @ 0.1\,Hz & V$_{\mathrm{N}}$ & 10\,nV$/\sqrt[]{\mathrm{Hz}}$\\
Voltage noise density @ 1\,Hz & V$_{\mathrm{N}}$ & 1.7\,nV$/\sqrt[]{\mathrm{Hz}}$\\
Voltage noise density @ 10\,Hz & V$_{\mathrm{N}}$ & 1\,nV$/\sqrt[]{\mathrm{Hz}}$\\
Voltage noise 0.1\,Hz to 10\,Hz\footnotemark[2] & V$_{\mathrm{N}}$ & 4.7\,nV$_{\mathrm{rms}}$\\

Bandwith (3\,dB) &  & 100\,kHz\\
CMRR\footnotemark[3] &  & $>$ 140\,dB\\
Common mode voltage range & V$_{\mathrm{P}}$, V$_{\mathrm{N}}$ & $\pm1$\,V\\
\hline
\end{tabular}
\footnotetext[1]{Measured at 10\,Hz.}
\footnotetext[2]{This range is typically used to compare 1/f noise of different amplifiers. Not convenient with this amplifier because white noise has big influence even below 10\,Hz.}
\footnotetext[3]{Common-mode rejection ratio, below 100\,Hz.}
\end{table}

\section{voltage-noise versus current-noise}
FETs with low voltage-noise typically have high input capacitance and high input leakage current. Therefore selecting an input FET is a trade-off between several parameters.
The selected FET IF3602 by InterFET is ideal for source impedances in the range of 1 M$\Omega$ and low-bandwidth use.
It is possible to employ the same approach with other FETs, suitable for higher source impedances or higher frequencies.

\section{conclusions}
Temperature variations have a big impact on the low-frequency noise of amplifiers. Stabilizing the temperature of the amplifier reduces low-frequency drifts dramatically.
To the best of our knowledge, no other room temperature amplifiers with similar performance have been realized by now.

The first prototypes have been used for the measurements presented in Ref.~\onlinecite{pauline}. The high contact resistance of the graphene samples required an amplifier with low leakage current.

The presented concepts are suitable to be used also in other labs and even commercial devices could employ our techniques.

\section{supplementary material}
See supplementary material for a more detailed schematic showing the current sources, iterative trimming of resistors, and response to large signal steps.

\begin{acknowledgments}
We would like to thank to A. Bosshard for the circuit board layout and E. Studer for the CAD drawings.
\end{acknowledgments}

\nocite{*}

\bibliography{dc_amp}

\providecommand{\noopsort}[1]{}\providecommand{\singleletter}[1]{#1}%
\begin{thebibliography}{11}%
\makeatletter
\providecommand \@ifxundefined [1]{%
 \@ifx{#1\undefined}
}%
\providecommand \@ifnum [1]{%
 \ifnum #1\expandafter \@firstoftwo
 \else \expandafter \@secondoftwo
 \fi
}%
\providecommand \@ifx [1]{%
 \ifx #1\expandafter \@firstoftwo
 \else \expandafter \@secondoftwo
 \fi
}%
\providecommand \natexlab [1]{#1}%
\providecommand \enquote  [1]{``#1''}%
\providecommand \bibnamefont  [1]{#1}%
\providecommand \bibfnamefont [1]{#1}%
\providecommand \citenamefont [1]{#1}%
\providecommand \href@noop [0]{\@secondoftwo}%
\providecommand \href [0]{\begingroup \@sanitize@url \@href}%
\providecommand \@href[1]{\@@startlink{#1}\@@href}%
\providecommand \@@href[1]{\endgroup#1\@@endlink}%
\providecommand \@sanitize@url [0]{\catcode `\\12\catcode `\$12\catcode
  `\&12\catcode `\#12\catcode `\^12\catcode `\_12\catcode `\%12\relax}%
\providecommand \@@startlink[1]{}%
\providecommand \@@endlink[0]{}%
\providecommand \url  [0]{\begingroup\@sanitize@url \@url }%
\providecommand \@url [1]{\endgroup\@href {#1}{\urlprefix }}%
\providecommand \urlprefix  [0]{URL }%
\providecommand \Eprint [0]{\href }%
\providecommand \doibase [0]{http://dx.doi.org/}%
\providecommand \selectlanguage [0]{\@gobble}%
\providecommand \bibinfo  [0]{\@secondoftwo}%
\providecommand \bibfield  [0]{\@secondoftwo}%
\providecommand \translation [1]{[#1]}%
\providecommand \BibitemOpen [0]{}%
\providecommand \bibitemStop [0]{}%
\providecommand \bibitemNoStop [0]{.\EOS\space}%
\providecommand \EOS [0]{\spacefactor3000\relax}%
\providecommand \BibitemShut  [1]{\csname bibitem#1\endcsname}%
\let\auto@bib@innerbib\@empty
\bibitem [{\citenamefont {Carlo}\ \emph {et~al.}(2006)\citenamefont {Carlo},
  \citenamefont {Zhang}, \citenamefont {McClure},\ and\ \citenamefont
  {Marcus}}]{2006}%
  \BibitemOpen
  \bibfield  {author} {\bibinfo {author} {\bibfnamefont {L.~D.}\ \bibnamefont
  {Carlo}}, \bibinfo {author} {\bibfnamefont {Y.}~\bibnamefont {Zhang}},
  \bibinfo {author} {\bibfnamefont {D.~T.}\ \bibnamefont {McClure}}, \ and\
  \bibinfo {author} {\bibfnamefont {C.~M.}\ \bibnamefont {Marcus}},\
  }\href@noop {} {\bibfield  {journal} {\bibinfo  {journal} {Rev. Sci.
  Instrum}\ }\textbf {\bibinfo {volume} {77}} (\bibinfo {year}
  {2006})}\BibitemShut {NoStop}%
\bibitem [{\citenamefont {Arnaboldi}\ \emph {et~al.}(2011)\citenamefont
  {Arnaboldi}, \citenamefont {Giachero}, \citenamefont {Gotti}, \citenamefont
  {Maino},\ and\ \citenamefont {Pessina}}]{2011}%
  \BibitemOpen
  \bibfield  {author} {\bibinfo {author} {\bibfnamefont {C.}~\bibnamefont
  {Arnaboldi}}, \bibinfo {author} {\bibfnamefont {A.}~\bibnamefont {Giachero}},
  \bibinfo {author} {\bibfnamefont {C.}~\bibnamefont {Gotti}}, \bibinfo
  {author} {\bibfnamefont {M.}~\bibnamefont {Maino}}, \ and\ \bibinfo {author}
  {\bibfnamefont {G.}~\bibnamefont {Pessina}},\ }\href {\doibase
  10.1109/TNS.2011.2171367} {\bibfield  {journal} {\bibinfo  {journal} {IEEE
  Transactions on Nuclear Science}\ }\textbf {\bibinfo {volume} {58}},\
  \bibinfo {pages} {3204} (\bibinfo {year} {2011})}\BibitemShut {NoStop}%
\bibitem [{\citenamefont {Karlquist}\ \emph {et~al.}(1997)\citenamefont
  {Karlquist}, \citenamefont {Cutler}, \citenamefont {Ingman}, \citenamefont
  {Johnson},\ and\ \citenamefont {Parisek}}]{ocxc}%
  \BibitemOpen
  \bibfield  {author} {\bibinfo {author} {\bibfnamefont {R.~K.}\ \bibnamefont
  {Karlquist}}, \bibinfo {author} {\bibfnamefont {L.~S.}\ \bibnamefont
  {Cutler}}, \bibinfo {author} {\bibfnamefont {E.~M.}\ \bibnamefont {Ingman}},
  \bibinfo {author} {\bibfnamefont {J.~L.}\ \bibnamefont {Johnson}}, \ and\
  \bibinfo {author} {\bibfnamefont {T.}~\bibnamefont {Parisek}},\ }in\ \href
  {\doibase 10.1109/FREQ.1997.639207} {\emph {\bibinfo {booktitle} {Proceedings
  of International Frequency Control Symposium}}}\ (\bibinfo {year} {1997})\
  pp.\ \bibinfo {pages} {898--908}\BibitemShut {NoStop}%
\bibitem [{\citenamefont {Alessandrello}\ \emph {et~al.}(1996)\citenamefont
  {Alessandrello}, \citenamefont {Brofferio}, \citenamefont {Bucci},
  \citenamefont {Camin}, \citenamefont {Cremonesi}, \citenamefont {Giuliani},
  \citenamefont {Nucciotti}, \citenamefont {Pavan}, \citenamefont {Pessina},
  \citenamefont {Previtali},\ and\ \citenamefont {Sablich}}]{1996}%
  \BibitemOpen
  \bibfield  {author} {\bibinfo {author} {\bibfnamefont {A.}~\bibnamefont
  {Alessandrello}}, \bibinfo {author} {\bibfnamefont {C.}~\bibnamefont
  {Brofferio}}, \bibinfo {author} {\bibfnamefont {C.}~\bibnamefont {Bucci}},
  \bibinfo {author} {\bibfnamefont {D.~V.}\ \bibnamefont {Camin}}, \bibinfo
  {author} {\bibfnamefont {O.}~\bibnamefont {Cremonesi}}, \bibinfo {author}
  {\bibfnamefont {A.}~\bibnamefont {Giuliani}}, \bibinfo {author}
  {\bibfnamefont {A.}~\bibnamefont {Nucciotti}}, \bibinfo {author}
  {\bibfnamefont {M.}~\bibnamefont {Pavan}}, \bibinfo {author} {\bibfnamefont
  {G.}~\bibnamefont {Pessina}}, \bibinfo {author} {\bibfnamefont
  {E.}~\bibnamefont {Previtali}}, \ and\ \bibinfo {author} {\bibfnamefont
  {G.}~\bibnamefont {Sablich}},\ }in\ \href {\doibase
  10.1109/NSSMIC.1996.591044} {\emph {\bibinfo {booktitle} {1996 IEEE Nuclear
  Science Symposium. Conference Record}}},\ Vol.~\bibinfo {volume} {1}\
  (\bibinfo {year} {1996})\ pp.\ \bibinfo {pages} {504--508 vol.1}\BibitemShut
  {NoStop}%
\bibitem [{\citenamefont {Paul~Horowitz}(1989)}]{bootstrapping}%
  \BibitemOpen
  \bibfield  {author} {\bibinfo {author} {\bibfnamefont {W.~H.}\ \bibnamefont
  {Paul~Horowitz}},\ }\href@noop {} {\emph {\bibinfo {title} {The Art of
  Electronics}}},\ \bibinfo {edition} {2nd}\ ed.\ (\bibinfo  {publisher}
  {Cambridge University Press},\ \bibinfo {year} {1989})\BibitemShut {NoStop}%
\bibitem [{lar()}]{largestepcoment}%
  \BibitemOpen
  \href@noop {} {}\bibinfo {note} {The response of the amplifier to large
  signal steps is described in the supplemental material.}\BibitemShut {Stop}%
\bibitem [{\citenamefont {{G. Scandurra, G. Cannat\`{a}, and C.
  Ciofi}}(2011)}]{scandurra}%
  \BibitemOpen
  \bibfield  {author} {\bibinfo {author} {\bibnamefont {{G. Scandurra, G.
  Cannat\`{a}, and C. Ciofi}}},\ }\href@noop {} {\bibfield  {journal} {\bibinfo
   {journal} {AIP Advances}\ }\textbf {\bibinfo {volume} {1}},\ \bibinfo
  {pages} {022144} (\bibinfo {year} {2011})}\BibitemShut {NoStop}%
\bibitem [{\citenamefont {{Keithley Instruments, Inc.}}(2016)}]{driven_guard}%
  \BibitemOpen
  \bibfield  {author} {\bibinfo {author} {\bibnamefont {{Keithley Instruments,
  Inc.}}},\ }\href@noop {} {\emph {\bibinfo {title} {Low Level Measurements
  Handbook - 7th Edition}}}\ (\bibinfo  {publisher} {Keithley},\ \bibinfo
  {year} {2016})\BibitemShut {NoStop}%
\bibitem [{\citenamefont {Allan}(1966)}]{Allan_1966}%
  \BibitemOpen
  \bibfield  {author} {\bibinfo {author} {\bibfnamefont {D.~W.}\ \bibnamefont
  {Allan}},\ }\href {\doibase 10.1109/PROC.1966.4634} {\bibfield  {journal}
  {\bibinfo  {journal} {Proceedings of the IEEE}\ }\textbf {\bibinfo {volume}
  {54}} (\bibinfo {year} {1966}),\ 10.1109/PROC.1966.4634}\BibitemShut
  {NoStop}%
\bibitem [{\citenamefont {Allan}(1987)}]{Allan_1987}%
  \BibitemOpen
  \bibfield  {author} {\bibinfo {author} {\bibfnamefont {D.~W.}\ \bibnamefont
  {Allan}},\ }\href@noop {} {\bibfield  {journal} {\bibinfo  {journal} {IEEE
  Transactions on Instrumentation and Measurement}\ }\textbf {\bibinfo {volume}
  {IM-36}} (\bibinfo {year} {1987})}\BibitemShut {NoStop}%
\bibitem [{\citenamefont {{P. Simonet, S. Hennel, H. Overweg, R. Steinacher, M.
  Eich, Y. Lee, R. Pisoni, P. M\"{a}rki, T. Ihn, M. Beck, J. Faist and K.
  Ensslin}}()}]{pauline}%
  \BibitemOpen
  \bibfield  {author} {\bibinfo {author} {\bibnamefont {{P. Simonet, S. Hennel,
  H. Overweg, R. Steinacher, M. Eich, Y. Lee, R. Pisoni, P. M\"{a}rki, T. Ihn,
  M. Beck, J. Faist and K. Ensslin}}},\ }\href@noop {} {\enquote {\bibinfo
  {title} {{Anomalous Coulomb drag between bilayer graphene and a GaAs electron
  gas}},}\ }\bibinfo {note} {(to be published)}\BibitemShut {NoStop}%
\end{thebibliography}%

\end{document}